\documentclass[prl,twocolumn,showpacs,superscriptaddress,nofootinbib]{revtex4-1}
\usepackage[T1]{fontenc} 
\usepackage{amsmath}
\usepackage{amssymb}
\usepackage{amsfonts}
\usepackage{graphicx}
\usepackage{enumitem}
\usepackage{bm}
\usepackage{rotating}
\usepackage{hyperref}
\usepackage{xcolor}
\usepackage{array}
\usepackage{lmodern}
\usepackage[normalem]{ulem}
\usepackage{braket}
\usepackage{tensor}
\usepackage{mathrsfs}
\usepackage{float}

\usepackage[normalem]{ulem}

\begin{document}

\title{Solving the strong CP problem by a $\bar{\theta}$-characterized mirror symmetry}

\author{Pei-Hong Gu}

\email{phgu@seu.edu.cn}

\affiliation{School of Physics, Jiulonghu Campus, Southeast University, Nanjing 211189, China}

\begin{abstract}

In the standard model QCD Lagrangian, a term of CP violating gluon density is theoretically expected to have a physical coefficient $\bar{\theta}$ of the order of unity. However, the upper bound on the electric dipole moment of neutron enforces the value of $\bar{\theta}$ to be extremely small. Such a huge gap between theoretical expectation and experimental result is commonly known as the strong CP problem. To solve this puzzle in an appealing context of two Higgs doublets, we propose an economical $\bar{\theta}$-characterized mirror symmetry between two Higgs singlets with respective discrete symmetries. In our scenario, the parameter $\bar{\theta}$ can completely disappear from the full Lagrangian after the standard model fermions take a proper phase rotation as well as the Higgs doublets and singlets. Moreover, all of new physics for solving the strong CP problem can be allowed near the TeV scale.

\end{abstract}


\maketitle

In the standard model QCD Lagrangian, a term of CP violating gluon density is theoretically expected to have a physical coefficient $\bar{\theta}$ of the order of unity. However, the upper bound on the electric dipole moment of neutron enforces the value of $\bar{\theta}$ to be extremely small. The huge gap between theoretical expectation and experimental result leads to the so-called strong CP problem \cite{kg2010,dgnv2020,sikive2021,workman2022}. In 1977, Peccei and Quinn found that the CP violating $\bar{\theta}$-term can be rotated away if the QCD Lagrangian contains a global symmetry \cite{pq1977}. This global symmetry now is well known as the Peccei-Quinn (PQ) symmetry $U(1)_{\textrm{PQ}}^{}$. After the PQ global symmetry is spontaneously broken, a massless Goldstone boson can emerge as usual. However, this Goldstone boson picks up a mass through the color anomaly \cite{adler1969,bj1969,ab1969} and hence becomes a pseudo Goldstone boson, named the axion \cite{weinberg1978,wilczek1978}.

The simplest approach to the PQ symmetry seems to consider a two-Higgs-doublet model \cite{pq1977}. Unfortunately, this original PQ model was quickly ruled out in experiments. In fact, the axion has not been observed experimentally and is still an invisible particle. This implies that the interactions between the axion and the SM particles should be at an extremely weak level \cite{kg2010,dgnv2020,sikive2021,workman2022}. For a successful realization of the PQ symmetry with an invisible axion, Kim-Shifman-Vainstein-Zakharov (KSVZ) \cite{kim1979,svz1980} and Dine-Fischler-Srednicki-Zhitnitsky (DFSZ) \cite{dfs1981,zhitnitsky1980} then proposed their elegant methods to effeciently decrease the couplings of the axion to the SM particles. Currently, the null results from axion searches strictly constrain the PQ symmetry in the KSVZ-type and DFSZ-type models \cite{kim1979,svz1980,dfs1981,zhitnitsky1980} to be spontaneously broken far far above the weak scale \cite{kg2010,dgnv2020,sikive2021,workman2022}.

In the KSVZ-type and DFSZ-type models, the new particles except the invisible axion should be too heavy to verify in experiments unless the related couplings are artificially small. This means all experimental attempts to test the PQ symmetry can only depend on the axion-meson mixing and hence the axion searches \cite{kg2010,dgnv2020,sikive2021,workman2022}. On the theoretical side, for a huge hierarchy between the PQ and electroweak symmetry breaking scales, the inevitable Higgs portal should have an extremely small coupling, otherwise, its contribution should have a large cancellation with the rarely quadratic term of the SM Higgs scalar \cite{cv2016}. In some sense, the invisible axion models pay a price of additional fine tuning to solve the strong CP problem.

In this paper we shall propose a new mechanism to solve the strong CP problem in the appealing context of two Higgs doublets \cite{bflrss2011}. Specifically, we shall introduce an economical $\bar{\theta}$-characterized mirror symmetry only between two Higgs singlets with respective discrete symmetries. In our scenario, the parameter $\bar{\theta}$ can completely disappear from the full Lagrangian after the standard model fermions and the Higgs scalars take a proper phase rotation. Moreover, all of new physics for solving the strong CP problem can be allowed near the TeV scale.

Before starting the details of our mechanism, permit us to briefly review the strong CP problem. The QCD Lagrangian of the SM is described as  
\begin{eqnarray}
\label{qcd}
\mathcal{L}_{\textrm{QCD}}^{}&=&\sum_{q}^{} \bar{q} \left(i D \!\!\!\!/ - m_q^{} e^{i\theta_q}_{} \right)q - \frac{1}{4} G^{a\mu\nu}_{} G_{\mu\nu}^{a} \nonumber\\
&& - \theta \frac{\alpha_s^{}}{8\pi} G^{a\mu\nu}_{}\tilde{G}_{\mu\nu}^{a}\,,
\end{eqnarray}
where $\theta_q^{}$ is the phase from the Yukawa couplings of quark fields, $\theta$ is the QCD vacuum angle,  $G^{a}_{\mu\nu}$ is the gluon field strength tensor and $\tilde{G}^{a}_{\mu\nu}$ is its dual. After the quark fields take the chiral phase transformation as below, 
\begin{eqnarray}
q\rightarrow e^{-i \gamma_5^{} \theta_q/2}_{}  q \,,
\end{eqnarray}
their mass terms can remove the phases $\theta_q^{}$ from the QCD Lagrangian, i.e. 
\begin{eqnarray}
\label{qcdphy}
\mathcal{L}_{\textrm{QCD}}^{}&=&\sum_{q}^{} \bar{q} \left(i D \!\!\!\!/ - m_q^{} \right)q - \frac{1}{4} G^{a\mu\nu}_{} G_{\mu\nu}^{a}  - \bar{\theta} \frac{\alpha_s^{}}{8\pi} G^{a\mu\nu}_{}\tilde{G}_{\mu\nu}^{a}\nonumber\\
&&\textrm{with}~~\bar{\theta}\equiv\theta- \textrm{Arg}\textrm{ Det} \left(M_d^{} M_u^{} \right)\,.
\end{eqnarray}
Here $M_{d}^{}$ and $M_{u}^{}$ are the respective mass matrices of the SM down-type and up-type quarks. To satisfy the upper limits on the electric dipole moment of neutron, the value of $\bar{\theta}$ should be extremely small rather than the theoretically expected order of unity, i.e.
\begin{eqnarray}
\left|\bar{\theta}\right|<10^{-10}_{}\,.
\end{eqnarray}
This fine tuning of ten orders of magnitude is commonly known as the strong CP problem.

\begin{table*}
\begin{center}
\begin{tabular}{|l||c|c|c|c||c|c|c|c|c|c|}  \hline &&&&&&&&&&\\[-2.0mm] ~~~$Scalars~\&~Fermions$~~~&~~~$\xi_1^{}$~~~&~~~$\xi_{2}^{}$~~~&~~~$\phi_{1}^{}$~~~&~~~$\phi_{2}^{}$~~~&~~~$q_{L}^{}$~~~& ~~~$d_{R}^{}$~~~&~~~$u_{R}^{}$~~~&~~~$l_{L}^{}$~~~&~~~$e_{R}^{}$~~~&~~~$N_{R}^{}$~~~\\
&&&&&&&&&&\\[-2.0mm]\hline&&&&&&&&&& \\[-1.5mm]~~~~~~~~~~~$SU(3)_c^{}$&$1$ &$1$ &$1$ &$1$ &$3$ &$3$ & $3$ & $1$&$1$ &$1$ \\
&&&&&&&&&&\\[-2.0mm]\hline&&&&&&&&&& \\[-1.5mm]~~~~~~~~~~~$SU(2)_L^{}$ &$1$ &$1$ &$1$ &$1$ &$2$ &$1$ & $1$ & $2$&$1$ &$1$  \\
&&&&&&&&&&\\[-2.0mm]\hline&&&&&&&&&& \\[-1.5mm]~~~~~~~~~~~~$U(1)_Y^{}$&$0$ &$0$ &$+\frac{1}{2}$ &$+\frac{1}{2}$ &$+\frac{1}{6}$ &$-\frac{1}{3}$ & $+\frac{2}{3}$ & $-\frac{1}{2}$&$-1$ &$0$  \\
&&&&&&&&&&\\[-2.0mm]\hline&&&&&&&&&& \\[-1.5mm]~~~~~~~~~~~~~$Z_{3}^{(1)}$&$e^{i\frac{2\pi}{3}}_{}$ &$1$ &$e^{i\frac{2\pi}{3}}_{}$ & $1$& $1$ &$e^{i\frac{4\pi}{3}}_{}$ & $1$ & $1$&$e^{i\frac{4\pi}{3}}_{}$ &$1$ \\
&&&&&&&&&&\\[-2.0mm]\hline&&&&&&&&&& \\[-1.5mm]~~~~~~~~~~~~~$Z_{3}^{(2)}$ &$1$&$e^{i\frac{2\pi}{3}}_{}$  & $e^{i\frac{2\pi}{3}}_{}$& $1$ &$1$ &$e^{i\frac{4\pi}{3}}_{}$ & $1$&$1$ &$e^{i\frac{4\pi}{3}}_{}$ &$1$\\
[1.5mm]
\hline
\end{tabular}
\vspace{0.25cm}
\caption{\label{fields} All of scalars and fermions in the model. The two Higgs singlets $\xi_{1,2}^{}$ are distinguished by the $Z^{(1)}_{3}\times Z^{(2)}_{3}$ discrete symmetries as well as the two Higgs doublets $\phi_{1,2}^{}$. Besides three generations of the SM quarks $q_L^{}$, $d_R^{}$ and $u_R^{}$, the SM leptons $l_L^{}$ and $e_R^{}$, we introduce three right-handed neutrinos $N_R^{}$ to realize a seesaw mechanism for the generation of tiny neutrino masses and also a leptogenesis mechanism for the explanation of cosmological baryon asymmetry. Here the family indices of the fermions are omitted for simplicity.}
\end{center}
\end{table*}

We now demonstrate our mechanism in a realistic model. All of scalars and fermions in the model are summarized in Table \ref{fields}. The two Higgs singlets $\xi_{1,2}^{}$ are distinguished by the $Z^{(1)}_{3}\times Z^{(2)}_{3}$ discrete symmetries as well as the two Higgs doublets $\phi_{1,2}^{}$. Besides three generations of the SM quarks $q_L^{}$, $d_R^{}$ and $u_R^{}$, the SM leptons $l_L^{}$ and $e_R^{}$, we introduce three right-handed neutrinos $N_R^{}$ to realize a seesaw \cite{minkowski1977,yanagida1979,grs1979,ms1980,sv1980} mechanism for the generation of tiny neutrino masses and also a leptogenesis \cite{fy1986} mechanism for the explanation of cosmological baryon asymmetry. Here the family indices of the fermions are omitted for simplicity. Moreover, there is a $\bar{\theta}$-characterized mirror symmetry between the two Higgs singlets $\xi_{1,2}^{}$, i.e.  
\begin{eqnarray}
\label{mirror}
\xi_{1}^{}\stackrel{\textrm{~$\bar{\theta}$-characterized mirror~symmetry }}{\leftarrow\!\!\!-\!\!\!-\!\!\!-\!\!\!-\!\!\!-\!\!\!-\!\!\!-\!\!\!-\!\!\!-\!\!\!-\!\!\!-\!\!\!-\!\!\!-\!\!\!-\!\!\!-\!\!\!-\!\!\!-\!\!\!-\!\!\!-\!\!\!-\!\!\!-\!\!\!-\!\!\!-\!\!\!-\!\!\!-\!\!\!\rightarrow} e^{-i\bar{\theta}/3}_{}\xi_{2}^{} \,.\end{eqnarray}
This may be a minimal version of the $\bar{\theta}$-characterized mirror symmetry \cite{gu2023}.

According to the charge assignments in Table \ref{fields}, we write down all of the allowed Yukawa and mass terms involving the fermions, i.e.
\begin{eqnarray}
\label{yukawa}
\mathcal{L}_{Y+M}^{}&=& -y_d^{}\bar{q}_L^{} \phi_1^{} d_R^{}  -y_u^{}\bar{q}_L^{} \tilde{\phi}_2^{} u_R^{}  -y_e^{}\bar{l}_L^{} \phi_1^{} e_R^{} \nonumber\\
[2mm]
&& -y_N^{}\bar{l}_L^{} \tilde{\phi}_2^{} N_R^{} -\frac{1}{2}M_N^{} \bar{N}_R^{} N_R^c +\textrm{H.c.} \nonumber\\
[2mm]
&&\textrm{with}~~\tilde{\phi}_{1,2}^{}=i\tau_2^{}\phi_{1,2}^\ast\,,
\end{eqnarray}
as well as the full scalar potential at renormalizable level, i.e. 
\begin{eqnarray}
\label{potential}
V&=& \mu_1^2 \phi_1^\dagger \phi^{}_1 + \mu_2^2 \phi^\dagger_2 \phi^{}_2  + \lambda_1^{} \left(\phi^\dagger_1 \phi^{}_1\right)^2_{}+ \lambda_2^{} \left(\phi^\dagger_2 \phi^{}_2\right)^2_{}
\nonumber\\
[2mm]
&&+ \lambda_3^{} \phi^\dagger_1 \phi^{}_1 \phi^\dagger_2 \phi^{}_2 
+\lambda_4^{} \phi^\dagger_1 \phi^{}_2 \phi^\dagger_2 \phi^{}_1+   \mu_\xi^2 \left(\xi_1^\ast \xi_1^{}+\xi_2^\ast\xi_2^{}\right) \nonumber\\
[2mm]
&&+\kappa_1^{}\left[\left(\xi_1^\ast \xi_1^{}\right)^2_{} + \left(\xi_2^\ast \xi_2^{}\right)^2_{}\right] +\kappa^{}_2  \xi_1^\ast \xi_1^{} \xi_2^\ast\xi_2^{}\nonumber\\
[2mm]
&&+\rho_\xi^{} \left[\left(\xi_1^3+e^{-i\bar{\theta}}_{}\xi_2^3 \right)+\textrm{H.c.}\right]+ \epsilon_1^{}\phi^\dagger_1 \phi^{}_1 \left(\xi^\ast_1\xi^{}_1 +\xi^\ast_2 \xi^{}_2\right) \nonumber\\
[2mm]
&&+ \epsilon_2^{}\phi^\dagger_2 \phi^{}_2 \left(\xi^\ast_1\xi^{}_1 +\xi^\ast_2 \xi^{}_2\right)+ \epsilon_3^{} \left(\xi_1^{} \xi_2^{} \phi^\dagger_1 \phi_2^{} +\textrm{H.c.}\right)\,.
\end{eqnarray}
Here the Yukawa couplings and the Majorana masses involving the right-handed neutrinos are responsible for the reaization of seesaw and leptogenesis. We will not study the details of seesaw and leptogenesis which are beyond the goal of the present work. It should be noted that the $\bar{\theta}$-characterized mirror symmetry (\ref{mirror}) is exactly complied in the full Lagrangian where the kinetic terms are not shown for simplicity.

We then clarify that the fields in Table \ref{fields} with the $\bar{\theta}$-characterized mirror symmetry in Eq. (\ref{mirror}) certainly can help us to solve the strong CP problem. Actually, after the two Higgs singlets, the two Higgs doublets and the three generations of fermions take the phase rotations as below,
\begin{eqnarray}
&&\left(\begin{array}{l}\xi_{1}^{}\rightarrow\xi_{1} \\
[3mm] \xi_{2}^{} \rightarrow   e^{+i\bar{\theta}/3}_{} \xi_2^{}\end{array} \right)\,,~~~~ \left(\begin{array}{l}
\phi_{1}^{}\rightarrow e^{+i\bar{\theta}/3}_{} \phi_{1}\\
[3mm] 
\phi_{2}^{}\rightarrow  \phi_{2}
\end{array}\right)\,,\nonumber\\
[3mm]
&&\left(\begin{array}{l}\,\,q_{L}^{}\rightarrow  q_{L}\\
[3mm]\,d_{R}^{}\rightarrow  e^{-i\bar{\theta}/3}_{}d_{R}\\
[3mm] 
\,u_{R}^{}\rightarrow  u_{R}\\
[3mm]
\,\,\,l_{L}^{} \rightarrow l_{L}\\
[3mm]
\,\,e_{R}^{}\rightarrow  e^{-i\bar{\theta}/3}_{}e_{R} \\
[3mm]
N_{R}^{} \rightarrow N_{R}^{}\\
\end{array}\right) \,,
\end{eqnarray}
the QCD Lagrangian (\ref{qcdphy}) and the scalar potential (\ref{potential}) can simultaneously remove the parameter $\bar{\theta}$, i.e. 
\begin{eqnarray}
\mathcal{L}_{\textrm{QCD}}^{} \!\!& \Rightarrow&\!\! \sum_{q}^{} \bar{q} \left(i D \!\!\!\!/ - m_q^{} \right)q - \frac{1}{4} G^{a\mu\nu}_{} G_{\mu\nu}^{a}\,,\\
[3mm]
\!\!\!\!\!\!\!\!\!\!\!\!V\!\!&\Rightarrow&\!\! \mu_1^2 \phi_1^\dagger \phi^{}_1 + \mu_2^2 \phi^\dagger_2 \phi^{}_2 + \lambda_1^{} \left(\phi^\dagger_1 \phi^{}_1\right)^2_{}+ \lambda_2^{} \left(\phi^\dagger_2 \phi^{}_2\right)^2_{}
\nonumber\\
[2mm]
\!\!\!\!\!\!\!\!\!\!\!\!\!\!&&\!\!+ \lambda_3^{} \phi^\dagger_1 \phi^{}_1 \phi^\dagger_2 \phi^{}_2 
+\lambda_4^{} \phi^\dagger_1 \phi^{}_2 \phi^\dagger_2 \phi^{}_1+   \mu_\xi^2 \left(\xi_1^\ast \xi_1^{}+\xi_2^\ast\xi_2^{}\right) \nonumber\\
[2mm]
\!\!\!\!\!\!\!\!\!\!\!\!\!\!&&\!\!+\kappa_1^{}\left[\left(\xi_1^\ast \xi_1^{}\right)^2_{} + \left(\xi_2^\ast \xi_2^{}\right)^2_{}\right] +\kappa^{}_2  \xi_1^\ast \xi_1^{} \xi_2^\ast\xi_2^{}\nonumber\\
[2mm]
\!\!\!\!\!\!\!\!\!\!\!\!\!\!&&\!\!+\rho_\xi^{} \left[\left(\xi_1^3+\xi_2^3 \right)+\textrm{H.c.}\right] + \epsilon_1^{}\phi^\dagger_1 \phi^{}_1 \left(\xi^\ast_1\xi^{}_1 +\xi^\ast_2 \xi^{}_2\right) \nonumber\\
[2mm]
\!\!\!\!\!\!\!\!\!\!\!\!\!\!&&\!\!+ \epsilon_2^{}\phi^\dagger_2 \phi^{}_2 \left(\xi^\ast_1\xi^{}_1 +\xi^\ast_2 \xi^{}_2\right)+ \epsilon_3^{} \left(\xi_1^{} \xi_2^{} \phi^\dagger_1 \phi_2^{} +\textrm{H.c.}\right)\,,\nonumber\\
\!\!\!\!\!\!\!\!\!\!\!\!\!\!&&\!\!
\end{eqnarray}
meanwhile, the Yukawa and mass terms (\ref{yukawa}) can keep invariant as well as the unshown kinetic terms.

When the Higgs scalars $\xi_{1}^{}$, $\xi_2^{}$, $\phi_1^{}$ and $\phi_{2}^{}$ respectively develop their nonzero vacuum expectation values $v_{\xi_1}^{}$, $v_{\xi_2}^{}$, $v_{\phi_1}^{}$ and $v_{\phi_2}^{}$, they can be expressed by  
\begin{eqnarray}
\xi_{1}^{} &=&\left(v_{\xi_1}^{}+h_{\xi_1}^{}+iP_{\xi_1}^{}\right)/\sqrt{2}\,,\\
[3mm]
\xi_{2}^{} &=&\left(v_{\xi_2}^{}+h_{\xi_2}^{}+iP_{\xi_2}^{}\right)/\sqrt{2}\,,\\
[3mm]
\phi_{1}^{}&=&\left[\begin{array}{c}
\phi^{+}_{1}\\
[2mm]
\left(v_{\phi_1}^{}+h_{\phi_1}^{}+i P_{\phi_1}^{}\right)/\sqrt{2}\end{array}\right]\,,\\
[3mm]
\phi_{2}^{}&=&\left[\begin{array}{c}
\phi^{+}_{2}\\
[2mm]
\left(v_{\phi_2}^{}+h_{\phi_2}^{}+i P_{\phi_2}^{}\right)/\sqrt{2}\end{array}\right]\,.
\end{eqnarray} 
Clearly, three would-be-Goldstone bosons, 
 \begin{eqnarray}
G_W^{\pm}&=&\left(v_{\phi_1}^{}\phi^{\pm}_{1} +v_{\phi_2}^{} \phi^{\pm}_{2} \right)/\sqrt{v_{\phi_1}^2 + v_{\phi_2}^2}\,,\\
[2mm]
G_Z^{}&=&\left(v_{\phi_1}^{}P^{}_{\phi_1} +v_{\phi_2}^{} P^{}_{\phi_2} \right)/\sqrt{v_{\phi_1}^2 + v_{\phi_2}^2}\,,
\end{eqnarray}
are eaten by the longitudinal components of the SM gauge bosons $W^{\pm}_{}$ and $Z$. Therefore, besides a pair of massive charged scalars,
\begin{eqnarray}
H^{\pm}_{}&=&\left(v_{\phi_1}^{}\phi^{\pm}_{2} - v_{\phi_2}^{} \phi^{\pm}_{1} \right)/\sqrt{v_{\phi_1}^2 + v_{\phi_2}^2}~~\textrm{with}\nonumber\\
[2mm]
&& m_{H^{\pm}}^2 = -\left[\lambda_4^{}+\epsilon_3^{} v_\xi^2 /\left(2 v_1^{} v_2^{}\right)\right]\left(v_1^2+v_2^2\right)\,,~~~~
\end{eqnarray}
we eventually obtain seven massive neutral scalars including four scalars and three pseudo scalars, i.e. 
\begin{eqnarray}
&&h_{\phi_1}^{}\,,~~h_{\phi_2}^{}\,,~~h_\xi^{}=\left(h^{}_{\xi_1} +h^{}_{\xi_2} \right)/\sqrt{2}\,,\nonumber\\
[2mm]
&&S_\xi^{} =\left(h^{}_{\xi_1} -h^{}_{\xi_2}\right)/\sqrt{2}\,;\\
[2mm]
&&a_\phi^{}=\left(v_{\phi_1}^{}P^{}_{\phi_2} -v_{\phi_2}^{} P^{}_{\phi_1} \right)/\sqrt{v_{\phi_1}^2 + v_{\phi_2}^2}\,,\nonumber\\
[2mm]
&&a_\xi^{}=\left(P^{}_{\xi_1} +P^{}_{\xi_2} \right)/\sqrt{2}\,,~~P_\xi^{}=\left(P^{}_{\xi_1} -P^{}_{\xi_2} \right)/\sqrt{2}\,.~~~~
\end{eqnarray}

With the minimum of the scalar potential, we have the mass-squared matrix of three scalars $h_{\phi_1}^{}$, $h_{\phi_2}^{}$ and $h_\xi^{}$, i.e. 
\begin{widetext}
\begin{eqnarray}
\label{matrixhiggs}
\!\!\!\!\!\!\!\!\!\!\!\!\mathcal{L}\supset-\frac{1}{2}\left[h_{\phi_1}^{}~h_{\phi_2}^{}~h_\xi^{}\right]\!\!\left[\begin{array}{ccc}
2\lambda_1^{}v_{\phi_1}^2 -\frac{1}{2}\epsilon_3^{} v_\xi^2 \frac{v_{\phi_2}^{}}{v_{\phi_1}^{}} & \left(\lambda_3^{}+\lambda_4^{}\right) v_{\phi_1}^{} v_{\phi_2}^{} +\frac{1}{2}\epsilon_3^{} v_\xi^2 & \sqrt{2} \left(\epsilon_1^{} +\frac{1}{2}\epsilon_3^{}\right) v_{\phi_1}^{} v_\xi^{}\\
[4mm]
 \left(\lambda_3^{}+\lambda_4^{}\right) v_{\phi_1}^{} v_{\phi_2}^{} +\frac{1}{2}\epsilon_3^{} v_\xi^2  & 2\lambda_2^{}v_{\phi_2}^2 -\frac{1}{2}\epsilon_3^{} v_\xi^2 \frac{v_{\phi_1}^{}}{v_{\phi_2}^{}} &  \sqrt{2} \left(\epsilon_2^{} +\frac{1}{2}\epsilon_3^{}\right) v_{\phi_2}^{} v_\xi^{} \\
 [4mm]
 \sqrt{2} \left(\epsilon_1^{} +\frac{1}{2}\epsilon_3^{}\right) v_{\phi_1}^{} v_\xi^{} &   \sqrt{2} \left(\epsilon_2^{} +\frac{1}{2}\epsilon_3^{}\right) v_{\phi_2}^{} v_\xi^{} & 2\kappa_1^{} v_\xi^2 + \frac{3}{\sqrt{2}}\rho_\xi^{} v_\xi^{} - \frac{1}{2} \epsilon_3^{} v_{\phi_1}^{} v_{\phi_2}^{} 
 \end{array}\right]\!\!\left[\begin{array}{c}h_{\phi_1}^{}\\
 [4mm]
 h_{\phi_2}^{}\\
 [4mm]
 h_\xi^{}\end{array}\right]\!.~
\end{eqnarray} 
\end{widetext}
Here and thereafter we have taken into account the fact,
\begin{eqnarray}
v_{\xi_1}^{}= v_{\xi_2}^{}\equiv v_{\xi}^{}\,,
\end{eqnarray}
which can be easily deduced from the minimization of the scalar potential. By diagonalizing the mass-squared matrix (\ref{matrixhiggs}), we can obtain three mass eigenstates $H_{1,2,3}^{}$ with Yukawa couplings. For simplicity, we do not perform this diagonalization in the present work. As for the forth scalar $S_\xi^{}$ without Yukawa couplings, it has been a mass eigenstate with the following mass square,
\begin{eqnarray}
m_{S_\xi}^2= 2\kappa_1^{} v_\xi^2 + \frac{3}{\sqrt{2}}\rho_\xi^{} v_\xi^{} - \frac{1}{2} \epsilon_3^{} v_{\phi_1}^{} v_{\phi_2}^{} \,.
\end{eqnarray}

We then consider the pseudo scalars $a_{\phi}^{}$, $a_{\xi}^{}$ and $P_\xi^{}$. Their mass-squared matrix is given by 
\begin{widetext}
\begin{eqnarray}
\mathcal{L}\supset-\frac{1}{2}\left[a_{\phi}^{}~~a_\xi^{}~~P_\xi^{}\right]\!\!\left[\begin{array}{ccc}
 -\frac{1}{2}\epsilon_3^{}v_\xi^2\left(\frac{v_{\phi_1}^{}}{v_{\phi_2}^{}} + \frac{v_{\phi_2}^{}}{v_{\phi_1}^{}}\right) & ~~-\frac{1}{4} \epsilon_3^{} v_\xi^{}\sqrt{v_{\phi_1}^2+v_{\phi_2}^2} &~~0\\
 [5mm]
 -\frac{1}{4} \epsilon_3^{} v_\xi^{}\sqrt{v_{\phi_1}^2+v_{\phi_2}^2}  &~~ -9\sqrt{2} \rho_\xi^{} v_\xi^{} -\epsilon_3^{} v_{\phi_1}^{} v_{\phi_2}^{} &~~0 \\
 [5mm]
 ~~0 &~~0 & ~~ -9\sqrt{2} \rho_\xi^{} v_\xi^{} -\epsilon_3^{} v_{\phi_1}^{} v_{\phi_2}^{} \end{array}\right]\!\!\left[\begin{array}{c}a_{\phi}^{}\\
 [5mm]
 a_\xi^{}\\
 [5mm]
 P_\xi^{}
 \end{array}\right].
\end{eqnarray} 
\end{widetext}
Clearly, $P_\xi^{}$ is already a mass eigenstate and its mass square is just 
\begin{eqnarray}
m_{P_\xi}^2= -9\sqrt{2} \rho_\xi^{} v_\xi^{} -\epsilon_3^{} v_{\phi_1}^{} v_{\phi_2}^{}  \,.
\end{eqnarray} 
As for $a_\phi^{}$ and $a_\xi^{}$, they mix with each other and their mass eigenstates are 
\begin{eqnarray}
a_1^{}&=& a_\phi^{} \cos\alpha - a_\xi^{} \sin\alpha~~\textrm{with}\nonumber\\
&&m_{a_1}^2=\frac{m_{a_\phi}^2 + m_{a_\xi}^2 + \sqrt{\left(m_{a_\phi}^2 - m_{a_\xi}^2 \right)^2_{}+4\Delta^4 }}{2}\,,\nonumber\\
&&\\
[3mm]
a_2^{}&=& a_\phi^{} \sin\alpha + a_\xi^{} \cos\alpha~~\textrm{with}\nonumber\\
&&m_{a_2}^2=\frac{m_{a_\phi}^2 + m_{a_\xi}^2 - \sqrt{\left(m_{a_\phi}^2 - m_{a_\xi}^2 \right)^2_{}+4\Delta^4 }}{2}\,,\nonumber\\
&&\,.
\end{eqnarray} 
Here $m_{a_\phi}^2$, $m_{a_\xi}^2$ and $\Delta^2_{}$ are defined by 
\begin{eqnarray}
m_{a_\phi}^2 &=&  -\frac{1}{2}\epsilon_3^{}v_\xi^2\left(\frac{v_{\phi_1}^{}}{v_{\phi_2}^{}} + \frac{v_{\phi_2}^{}}{v_{\phi_1}^{}}\right) \,,\nonumber\\
[3mm]
m_{a_\xi}^2 &=&-9\sqrt{2} \rho_\xi^{} v_\xi^{} -\epsilon_3^{} v_{\phi_1}^{} v_{\phi_2}^{}\,,\nonumber\\
[3mm]
\Delta^2&=& -\frac{1}{4} \epsilon_3^{} v_\xi^{}\sqrt{v_{\phi_1}^2+v_{\phi_2}^2}\,,
\end{eqnarray} 
while $\alpha$ is the mixing angle and is determined by 
\begin{eqnarray}
\tan 2\alpha = \frac{2\Delta^2_{}}{m_{A_\phi}^2 - m_{A_\xi}^2} \,.
\end{eqnarray} 
Since the pseudo scalars $a_{1,2}^{}$ couple to the axial currents of the SM quarks, they indeed act as the role of heavy axions \cite{gu2023}.

In the present paper, we have proposed a novel $\bar{\theta}$-characterized mirror symmetry to naturally solve the strong CP problem. In our scenario, the scalars include two Higgs singlets and two Higgs doublets, while the fermions include three generations of the SM fermions and the right-handed neutrinos. The $\bar{\theta}$-characterized mirror symmetry is only involved in the two Higgs singlets with respective discrete symmetries. The parameter $\bar{\theta}$ can completely disappear from the full Lagrangian after the fermions and the Higgs scalars take a proper phase rotation. Our mechanism allows that the new physics for solving the strong CP problem can be near the TeV scale.

\textbf{Acknowledgement}: This work was supported in part by the National Natural Science Foundation of China under Grant No. 12175038.

\end{document}